\documentclass[
  aps,
  prb,
  reprint,
  superscriptaddress,
  longbibliography,
  showkeys,
  citeautoscript,
]{revtex4-1}
\usepackage[intlimits]{amsmath}
\usepackage{amssymb}
\usepackage{amsthm}
\usepackage{booktabs}
\usepackage{braket}
\usepackage{color}
\usepackage{float}
\usepackage{graphicx}
\usepackage{glossaries}
            
\usepackage[colorlinks=true,
            linkcolor=black,
            citecolor=black,
            filecolor=black,
            urlcolor=black
            ]{hyperref}
\usepackage[utf8]{inputenc}
\usepackage[version=4]{mhchem}
\usepackage{multirow}
\usepackage{soul}
\usepackage{tabularx}
\usepackage{units}
\usepackage[dvipsnames]{xcolor}

\usepackage{dcolumn}
\usepackage{float}
\usepackage{booktabs}

\setacronymstyle{long-short}
\newacronym{qd}{QD}{quantum dot}
\newacronym{pl}{PL}{Photoluminescence}
\newacronym{cl}{CL}{Cathodoluminescence}
\newacronym{gaas}{GaAs}{Gallium Arsenide}
\newacronym{ingaas}{InGaAs}{Indium Gallium Arsenide}
\newacronym{ebl}{EBL}{Electron beam lithography}

\newacronym{vis}{VIS}{visible}
\newacronym{nir}{NIR}{near-infrared}
\newacronym{uv}{UV}{ultraviolet}
\newacronym{mse}{MSE}{mean squared error}
\newacronym{si}{SI}{Supporting Information}

\newcommand{\dtu}{
    Department of Electrical and Photonics Engineering, Technical University of Denmark,
    2800 Kgs. Lyngby, Denmark
}

\begin{document}

\title{Tailoring Polarization in WSe$_2$ Quantum Emitters through Deterministic Strain Engineering}

\author{Athanasios Paralikis}
\affiliation{\dtu}

\author{Claudia Piccinini}
\affiliation{\dtu}

\author{Abdulmalik A. Madigawa}
\affiliation{\dtu}

\author{Pietro Metuh}
\affiliation{\dtu}

\author{Luca Vannucci}
\affiliation{\dtu}

\author{Niels Gregersen}
\affiliation{\dtu}

\author{Battulga Munkhbat}
\email[]{bamunk@dtu.dk}
\affiliation{\dtu}

\begin{abstract}

Quantum emitters in transition metal dichalcogenides (TMDs) have recently emerged as a promising platform for generating single photons for optical quantum information processing. In this work, we present an approach for deterministically controlling the polarization of fabricated quantum emitters in a tungsten diselenide (WSe$_2$) monolayer. We employ novel nanopillar geometries with long and sharp tips to induce a controlled directional strain in the monolayer, and we report on fabricated WSe$_2$ emitters producing single photons with a high degree of polarization $(99\pm 4 \%)$ and high purity ($g^{(2)}(0) = 0.030 \pm 0.025$). Our work paves the way for the deterministic integration of TMD-based quantum emitters for future photonic quantum technologies.

\end{abstract}


\maketitle

\section{Introduction}

Within photonic quantum information technologies, a key component is the source of single photons \cite{gregersen_mørk_2017} used to encode the quantum bits. The ideal single-photon source (SPS) should feature deterministic emission of single indistinguishable photons with near-unity collection efficiency and high purity. The spontaneous parametric down-conversion process is a straightforward method \cite{Kwiat1995} for producing highly indistinguishable photons and has been the workhorse of the community for several decades; however, its probabilistic nature results in a significant trade-off between photon purity and efficiency. As an alternative, the two-level system in a semiconductor host material \cite{Aharonovich2016Solid-stateEmitters} has recently emerged as an attractive platform for the deterministic generation of single photons using the spontaneous emission process. By placing a semiconductor quantum dot (QD) in an optical cavity \cite{heindel2023quantum} and by exploiting cavity quantum electrodynamics (cQED) to control the light emission, emission of pure highly indistinguishable photons have been generated with efficiency as high as $\sim$ 0.6 \cite{tomm2021bright,Wang2019}. However, the fabrication of high-quality semiconductor QDs requires expensive, complex molecular beam epitaxy growth methods, which represent a significant drawback in developing QD-based sources.

As an alternative, quantum emitters in transition metal dichalcogenides (TMDs) are attractive due to their availability, their ease of integration into nanophotonic structures, and the versatility in engineering their photonic characteristics \cite{Montblanch2023LayeredTechnologies}. While the direct bandgap structure \cite{Le_2015, Montblanch2023LayeredTechnologies} of the pristine monolayer form of TMDs allows for an efficient classical light generation, the emission of pure single photons requires additional engineering. The microscopic origin of single-photon emission in TMDs is believed to be from localized excitonic states appearing either due to defects in the crystal lattice or due to the local strain\cite{Aharonovich2016Solid-stateEmitters, Gao2023Atomically-thinCommunication, Iff2019Strain-TunableMonolayers, Linhart2019LocalizedWSe2, Parto2021DefectK, Branny2017DeterministicSemiconductor}. Thus, the implementation of single-photon emitters (SPEs) into TMDs has been pursued by using deterministic defect fabrication \cite{Parto2021DefectK, xu2023conversion, wang2022utilizing} as well as by introducing localized strain in TMD monolayers \cite{Kumar2015Strain-InducedWSe2, Kern2016NanoscaleWSe2}. While stress can be applied directly to the flat TMD lattice \cite{Iff2019Strain-TunableMonolayers, So2021PolarizationWSe2, xu2023conversion, yu2021site}, strain is typically introduced by placing the TMD layer on arrays of nanopillars \cite{Parto2021DefectK, Iff2021Purcell-EnhancedCavity, Branny2017DeterministicSemiconductor, Palacios-Berraquero2017Large-scaleSemiconductors, kim2019position, zhao2021site, Luo2018DeterministicNanocavities, AzzamPartoMoody+2023+477+484, Kern2016NanoscaleWSe2} leading to strain concentration at the contact point. 

Similar to QDs, efficient single-photon emission can be obtained from an SPE in a TMD by placing it inside an optical cavity and exploiting cQED. A WSe$_2$ emitter was recently placed inside an open cavity geometry \cite{Drawer2023Monolayer-BasedCoherence} leading to the demonstration of an efficiency of $\sim$ 0.65 and an indistinguishability of $\sim$ 0.02. 
The cQED effect relies on a matching of polarizations of the dipole of the emitter and the electric field of the optical cavity mode. 
For the strain-induced SPE, a deformation strain potential---a nanowrinkle, serving as a macroscopic host for the SPE, is produced by the contact of the WSe$_2$ monolayer with the nanopillar. The polarization of the SPE dipole is then aligned \cite{So2021PolarizationWSe2, AzzamPartoMoody+2023+477+484, wang2021highly} with the orientation of the wrinkle. Control of the lateral dipole orientation is unnecessary in a rotationally symmetric geometry \cite{Drawer2023Monolayer-BasedCoherence} with two degenerate orthogonally polarized optical cavity modes, where the dipole can couple into either mode.
However, the general case requires control of the SPE dipole polarization, and thus the wrinkle orientation, to efficiently couple light into, for example, the polarized TE mode of a planar ridge waveguide as needed for on-chip integrated quantum photonics \cite{Wang2020}. 

In this work, we propose an efficient method for fabricating orientation-controlled nanowrinkles in monolayer WSe$_2$, and we demonstrate single photon emission with high purity ($g^{(2)}(0) = 0.030 \pm 0.025$) and a high degree of polarization (99 $\pm$ 4\%). This method consists of transferring flakes onto a nanostructure with long and sharp tips to form one-dimensional (1D) nanowrinkles, thereby controlling the direction of 1D strain and the polarization of the emitted photons. This work paves the way for integrating TMD-based quantum emitters into photonic systems.


\section{Results and Discussion}

\subsection{WSe$_2$ quantum emitters via cylindrical nanopillars}

\begin{figure}[t!]
\includegraphics[width=\columnwidth]{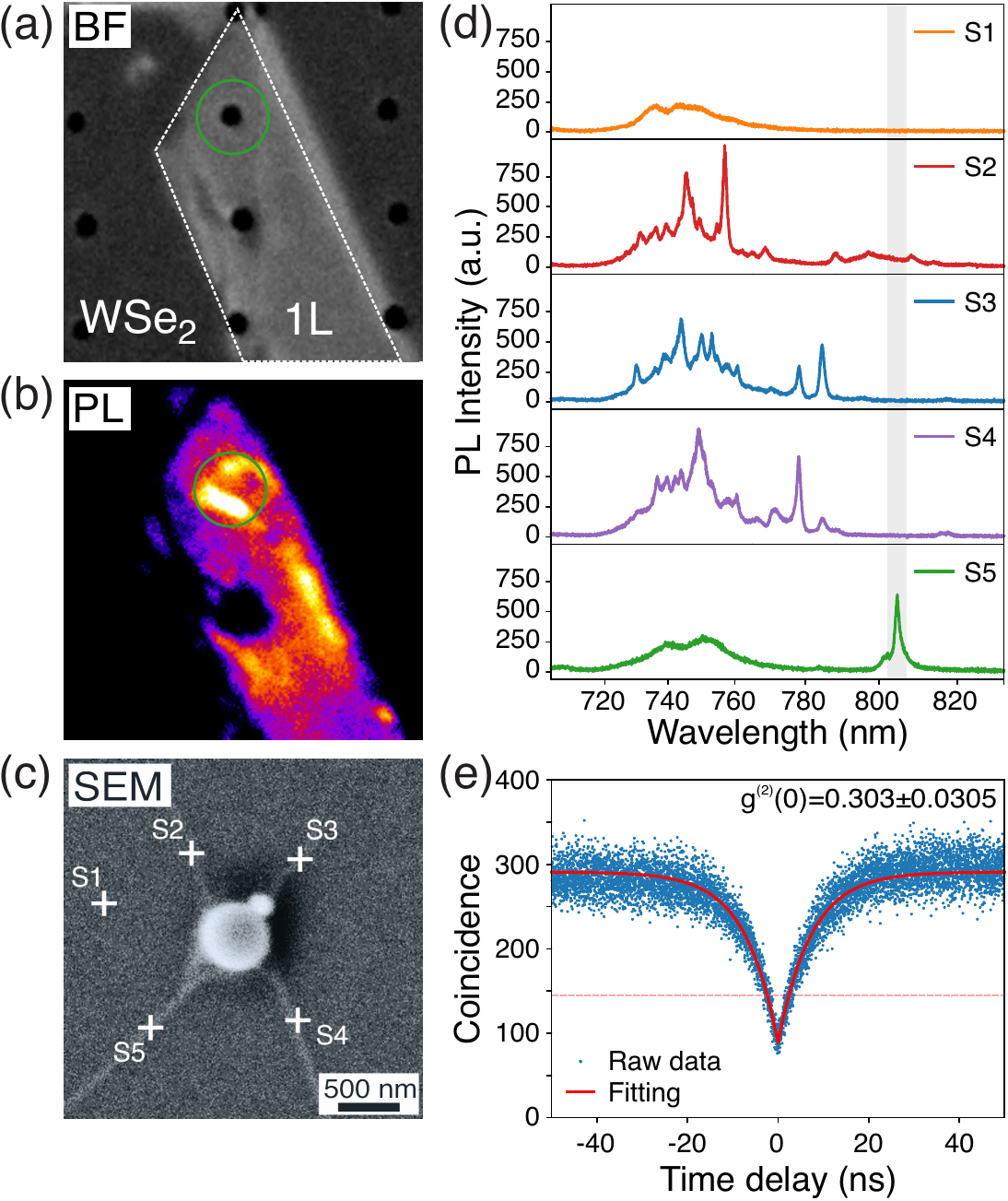}
\caption{\textbf{WSe\textsubscript{2} quantum emitters with cylindrical pillars.} (a) Bright-field (BF) image of monolayer WSe\textsubscript{2} deposited on cylindrical nanopillars on a Si/SiO\textsubscript{2} substrate. The white outline indicates the WSe$_2$ monolayer region. (b) Photoluminescence (PL) image of the WSe\textsubscript{2} monolayer flake taken at T = 4 K. The green circle indicates the region around the nanopillar. (c) SEM image of the nanopillar region circled in (a) and (b), revealing four nanowrinkles that are formed WSe\textsubscript{2} around the pillar. These nanowrinkles are likely to host the potential quantum emitters due to the strain. The crosses indicate the centers of the collection spots for each PL spectra presented in (d). (d) PL spectra collected from WSe\textsubscript{2} sample at T = 4 K around the five crosses depicted in (c). The obtained PL spectra from S2-5 show narrow emission lines, in contrast to a broad PL emission signal collected from the planar region (S1).
(e) Exemplary second-order correlation measurement ($g^{(2)}(\tau)$) of an isolated peak on the presented sample (S5), resulting in $g^{(2)}(0) = 0.303\pm0.035$.}
\label{Figure 1}
\end{figure}

\begin{figure*}[t!]
\includegraphics[width=0.9 \textwidth]{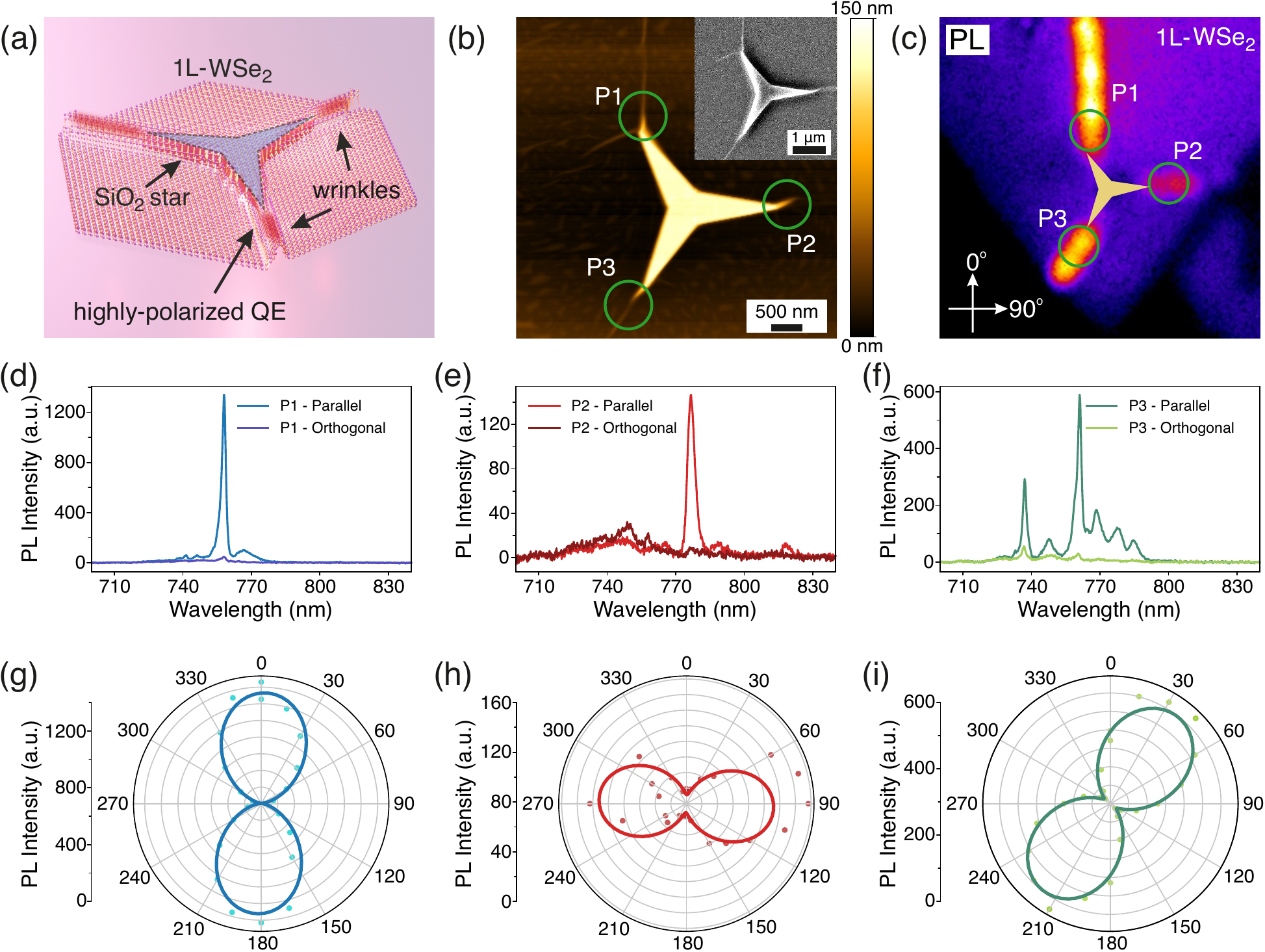}
\caption{\textbf{Highly polarized emission from WSe\textsubscript{2} quantum emitters with three-pointed star (TS) shaped pillars.} (a) A sketch of monolayer-based WSe\textsubscript{2} quantum emitters with a star-shaped nanopillar design with a high aspect ratio. (b) AFM image of a WSe\textsubscript{2} monolayer flake deposited on the TS nanostructure depicted in (a). The inset in (b) shows the corresponding SEM image. Both images reveal the nanowrinkles forming along the vertices of the TS. 
The directionality of the formed nanowrinkles mostly follows the directions of the vertices. 
(c) Corresponding PL image taken at T = 4 K of the fabricated sample exhibits brighter localized emission signals along the nanowrinkles, in contrast to the planar region. The two orthogonal axes are given as a reference for the subsequent polarization-resolved spectra and polar plots. In both (b) and (c), the three collection regions (P1-3) are highlighted.
(d-f) In-plane polarization-resolved PL spectra taken from three highlighted regions (P1-3), respectively. The presented PL spectra in (d-f) are measured under parallel and orthogonal in-plane polarization with respect to the direction of each nanowrinkle.
(g-i) Corresponding polar plots of the polarization-resolved PL intensities for each peak, color-coded to match the graphs in (d-f). The solid lines represent sinusoidal fits \cite{koudinov2004optical} of the data, revealing a degree of linear polarization of 99 ± 4\%, 82 ± 14\%, and 93 ± 3\%, respectively.}
\label{Figure 2}
\end{figure*}

We initially fabricate SPEs by transferring monolayer WSe$_2$ flakes on an array of conventional, \cite{Parto2021DefectK, Palacios-Berraquero2017Large-scaleSemiconductors} rotationally symmetric nanopillars.
The pillar dimensions were set to $\sim$ 150 nm in height and 500 nm in diameter. During the transfer, the monolayer adapts to the shape of the pillar due to the application of mechanical strain and is deformed by creating nanowrinkles.
The detailed fabrication process is discussed in the Methods section and Figure S1 and Supporting Note 1. In addition to strain engineering, we introduce defects to the monolayer WSe$_2$ via e-beam irradiation. Such a process is motivated in Supporting Note 2 and Figure S2 of the Supporting Information.
Figure 1a shows a bright-field (BF) image of the fabricated sample with cylindrical pillars, where the WSe$_2$ monolayer is highlighted with a white outline, and the pillars are identified through their contrast difference, appearing as dark spots. Using a 470 nm light-emitting diode (LED) as an excitation source, we perform photoluminescence (PL, T = 4 K) imaging on the same sample and present it in Figure 1b. We observe increased photon emission in proximity of a nanopillar (green circle) compared to the emission in the planar region of WSe$_2$.
This PL increase could be attributed to the efficient migration of excitons toward lower-energy states in the strained areas. Interestingly, the brighter PL emission signal seems to be located around the pillar. This suggests that a stronger strain is localized around the pillar, not on top, which is in good agreement with previously reported results\cite{Branny2017DeterministicSemiconductor}. However, the PL image does not clarify how the nanopillar induces local strain in the monolayer. \\
To provide microscopic details of the deformation around the pillar, we performed scanning electron microscope (SEM) imaging of the fabricated sample, presented in Figure 1c. The obtained SEM image reveals the formation of four straight nanowrinkles around the pillar.
It is generally accepted that these nanowrinkles host quantum emitters due to local strain, and their orientation is what dictates the polarization profile of the emitted single photons from the resulting emitters \cite{So2021PolarizationWSe2, AzzamPartoMoody+2023+477+484, wang2021highly}.

Due to the rotational symmetry of the nanopillar, which applies strain uniformly, the final number and orientation of these nanowrinkles are random, rendering us unable to control these attributes. Moreover, their proximity could affect further optical characterizations, such as purity.\\
To investigate the impact of closely formed nanowrinkles in the presence of a single pillar on the optical properties of the sample, we measured PL spectra (T = 4 K) from five different spots around the pillar, highlighted with white crosses (S1-S5) and present them in Figure \ref{Figure 1}d. A broad emission signal is observed from the planar monolayer region (S1) due to the delocalized exciton and trion states, whereas the collected spectra taken from different spots around the nanowrinkles (S2-S5) exhibit new discrete and narrow emission lines with higher intensities ranging from 730 to 810 nm \cite{Palacios-Berraquero2017Large-scaleSemiconductors}. These narrow emission lines are attributed to the new localized excitons hosted in the confined systems produced by the strain and defects in the mono- and few-layer WSe$_2$ \cite{Montblanch2023LayeredTechnologies, dang2020identifying, xu2023conversion, Linhart2019LocalizedWSe2, Parto2021DefectK,Kern2016NanoscaleWSe2,iff2018deterministic,Errando-Herranz2021ResonanceEmitters}.

Due to the collection area being larger than the emitter region, we partially collect light from the planar monolayer region, resulting in the broad emission being observed in the spectra from the nanowrinkles (S2-S5) ranging between 720-770 nm.
This makes isolating individual emission lines associated with potential single-photon emission from the broad multiphoton emission background relatively challenging, even with narrow optical band-pass filters. Moreover, due to the spatial proximity of the nanowrinkles, the PL spectra exhibit overlapping emission lines as we capture light emitted from multiple emitters simultaneously. These emission lines from different nanowrinkles oriented in the same directions might possess similar polarization profiles (S2-S4, S3-S5), further complicating the process of isolating individual emission lines with polarization filtering. Nevertheless, individual emission lines above 780 nm can be isolated using a long-pass or narrow band-pass optical filter (e.g., S5 in Figure \ref{Figure 1}d). 
 
To evaluate the single-photon purity of these emission lines, we performed the second-order autocorrelation ($g^{(2)}(\tau)$) measurement from a representative emission line ($\lambda \approx 807$ nm) under continuous-wave (CW) excitation at 455 nm via a Hanbury Brown–Twiss (HBT) setup. In this setup, the emitted photons are coupled to a 50:50 fiber beam splitter, whose output signals are detected by superconducting nanowire single-photon detectors (SNSPDs). The emission line at 807 nm was spectrally isolated using a 750 nm long-pass filter. Figure \ref{Figure 1}e shows the obtained correlation histogram, exhibiting a clear signature of photon antibunching. The extracted value of $g^{(2)}(0) = 0.303 \pm0.035$ is below the $g^{(2)}(0)<0.5$ threshold, indicating single-photon emission behavior. The reduced purity obtained from the particular emitter can be attributed to emission leakage from nearby SPEs that fall within the diffraction limit of our collection spot. This could be overcome by spatially isolating the nanowrinkles via novel nanostructure designs and subsequently controlling their directionality.   
The challenges posed by the lack of control of the localization and polarization in SPEs based on small cylindrical nanopillars emphasize the need for an effective strain-engineering approach to overcome these limitations.


\subsection{Control of nanowrinkle orientation via novel nanopillar designs}

We propose tailored nanostructure designs consisting of polygonal structures with long and sharp tips. As illustrated in Figure \ref{Figure 2}a, their shape allows strain to be induced by the vertices, facilitating the creation of spatially isolated nanowrinkles. 
Commanding the directionality of the one-dimensional SPE-hosting nanowrinkles gives rise to an effective control of the polarization of the emitted single photons \cite{So2021PolarizationWSe2, AzzamPartoMoody+2023+477+484, wang2021highly}.
Here, we explore three different geometries, namely, a three-pointed star (TS), a five-pointed star (FS), and a bowtie (BT). The structures are patterned out of negative e-beam resist (HSQ), which is cured into SiO$_2$ upon e-beam exposure. The thickness of the nanostructures ($\sim 150$ nm) was identified as optimal for producing nanowrinkles during the transfer process. Further details of the e-beam lithography and the remaining fabrication process are described in the Methods section. 

First, we demonstrate WSe$_2$ single-photon emitters with a TS nanostructure. Figure \ref{Figure 2}a shows a 3D schematic representation of the TS nanostructure with a WSe$_2$ monolayer flake, illustrating the formation of nanowrinkles that follow the directionality of each arm. The surface morphology of a sample hosting a WSe$_2$ monolayer flake is shown in the atomic force microscope (AFM) image in Figure \ref{Figure 2}b. The false coloring highlights the fabricated nanowrinkles versus the planar monolayer, with the inset providing a high-resolution SEM image of the same structure for clarity. We examine the morphological profiles of the fabricated nanowrinkles on the different designs in Figures S3,4,5 of the Supporting Information. We notice clear similarities in the nanowrinkles between the different geometries, meaning that a similar amount of strain is applied to the monolayer WSe$_2$ regardless of the design. More importantly, the nanowrinkles seem to be forming at the vertices, which was one of the primary reasons for tailoring these designs. Moreover, due to the size of the structures, the nanowrinkles are separated by more than 2 \textmu m, allowing us to address each one individually. Nevertheless, the alignment between the arms and the nanowrinkles is imperfect and will be addressed in the following sections.

To investigate how these nanowrinkles modify the optical properties of monolayer WSe$_2$, we performed PL imaging under a 470 nm LED excitation in a closed-cycle optical cryostat equipped with a microscope objective (60$\times$, NA = 0.82) operating at T = 4 K.
The obtained PL image (T = 4 K) taken from the TS sample is presented in Figure \ref{Figure 2}c, where increased PL emission is observed along the nanowrinkles. This increased PL emission is a sign of the successful fabrication of SPEs. 
Interestingly, the emission from the top of the TS structure, highlighted by a semi-transparent star, is comparable to that of the planar monolayer. This observation suggests that the mechanical strain applied at the sides of the arms or the top of the structure may not alter the bands significantly. In addition, forming a nanowrinkle seems crucial to induce the confinement effect, ultimately resulting in the generation of SPEs. Moreover, the increased PL signal is not only focused on the tips of the arms but can also be seen across the length of the nanowrinkles.
To gain some insights into the effect of the formed nanowrinkles on the polarization of the emitted photons, we performed the polarization-resolved \textmu PL measurement by mounting a linear polarizer to the characterization setup presented in the Methods section. 
The collected PL spectra from the individual nanowrinkles (P1-P3) under parallel and orthogonal polarizations with respect to the directionality of each nanowrinkle are presented in Figures \ref{Figure 2}d-f. All three emitters exhibit highly polarized emission, as the PL intensity of the orthogonal polarization is orders of magnitude lower than the parallel one.
To gain a better understanding, we analyzed the polarization-resolved \textmu PL spectra taken from the individual nanowrinkles. The extracted PL intensity as a function of the in-plane polarization exhibits highly linear polarized behavior, as shown in Figure \ref{Figure 2}g-i. To quantify the degree of linear polarization (DOLP) of the emitted photons, we use a sinusoidal fitting function and extract the DOLP using the maximum and minimum PL intensities as $(I_\text{max} - I_\text{min})/(I_\text{max} + I_\text{min})$, with the calculated values of 99 ± 4\%, 82 ± 14\%, and 93 ± 3\%, respectively for the three emitters. The strong degree of linear polarization is attributed to the nanowrinkles, which give rise to a localized one-dimensional strain potential, thus forming strongly aligned dipoles. \cite{tripathi2018spontaneous, Drawer2023Monolayer-BasedCoherence} In addition, the fitting reveals that the in-plane polarization profile of each emitter does follow the directionality of their respective nanowrinkles.
Hence, the proposed design facilitates the successful fabrication of spatially isolated SPEs in monolayer WSe$_2$. Despite the effective fabrication of SPEs, the TS structures produce nanowrinkles that are not perfectly aligned with their respective arms.
With the purpose of reducing the extension of the nanowrinkles to create spatially confined emitters, we introduce the FS structure. We speculate that by increasing the number of vertices while maintaining the size of the structures, the nanowrinkles will shorten and be more localized around the tips.

\begin{figure}[t!]
\includegraphics[width=\columnwidth]{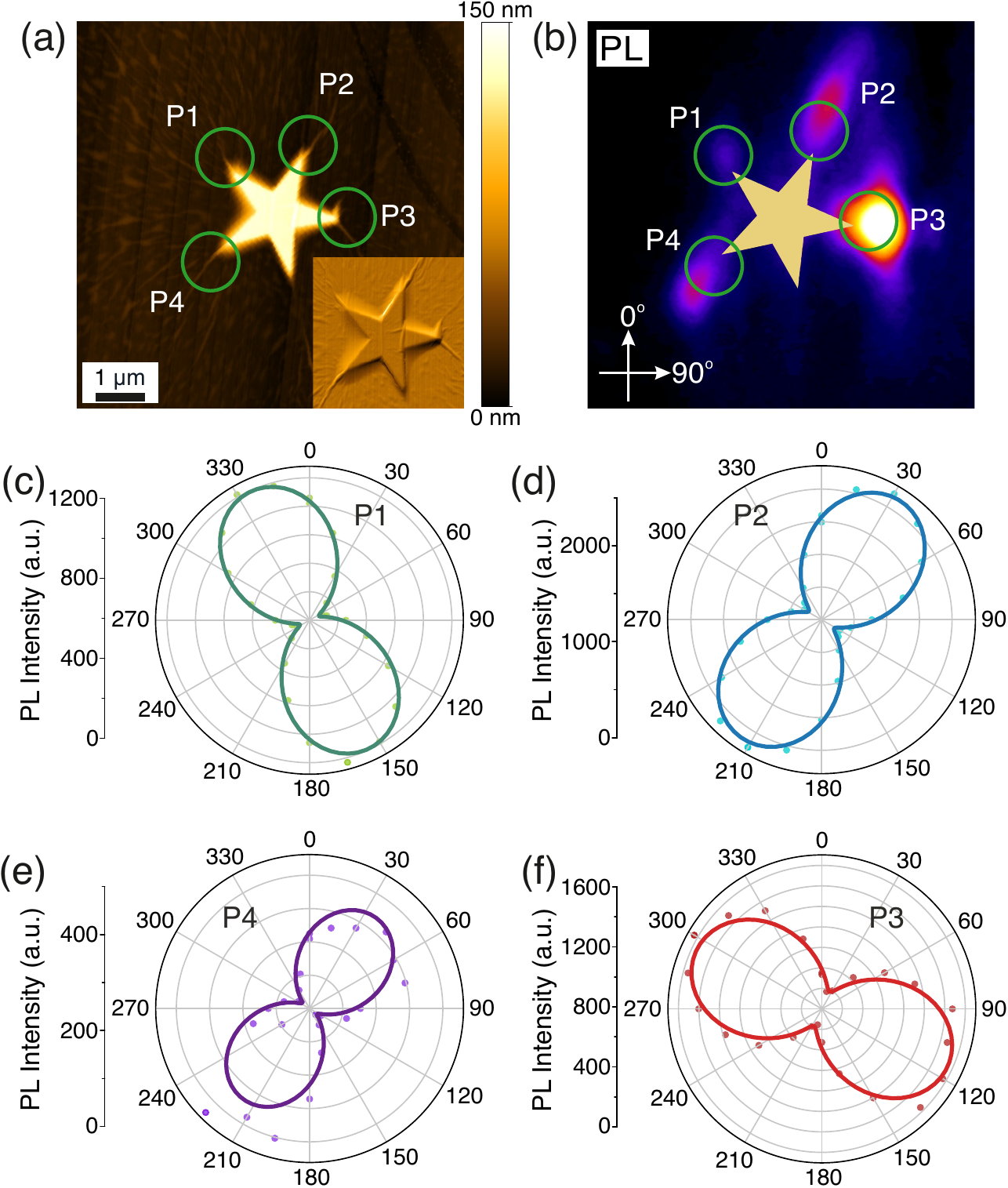}
\caption{\textbf{WSe\textsubscript{2} quantum emitters with a five-pointed star (FS) shaped pillars.} (a) AFM image of a WSe\textsubscript{2} monolayer flake deposited on the FS nanostructure and false-shading representation (inset). Both the main image and the inset reveal nanowrinkles forming at the vertices, mostly following their directionality. (b) The corresponding PL image at T = 4 K reveals increased photon emission along the nanowrinkles. The two orthogonal axes are given as a reference for the upcoming polarization-resolved polar plots. Both (a) and (b) highlight the collection spots (P1-4) for the polarization-resolved polar plots to follow.
 (c-f) Corresponding polar plots of the polarization-resolved PL intensities collected from P1-4, fitted with a sinusoidal function. The calculated degree of linear polarization is $86 \pm 2\%$, $81 \pm 2\%$, $76 \pm 3\%$, and $85 \pm 11\%$ for each peak respectively.}
\label{Figure 3}
\end{figure}

\begin{figure}[t!]
\includegraphics[width=\columnwidth]{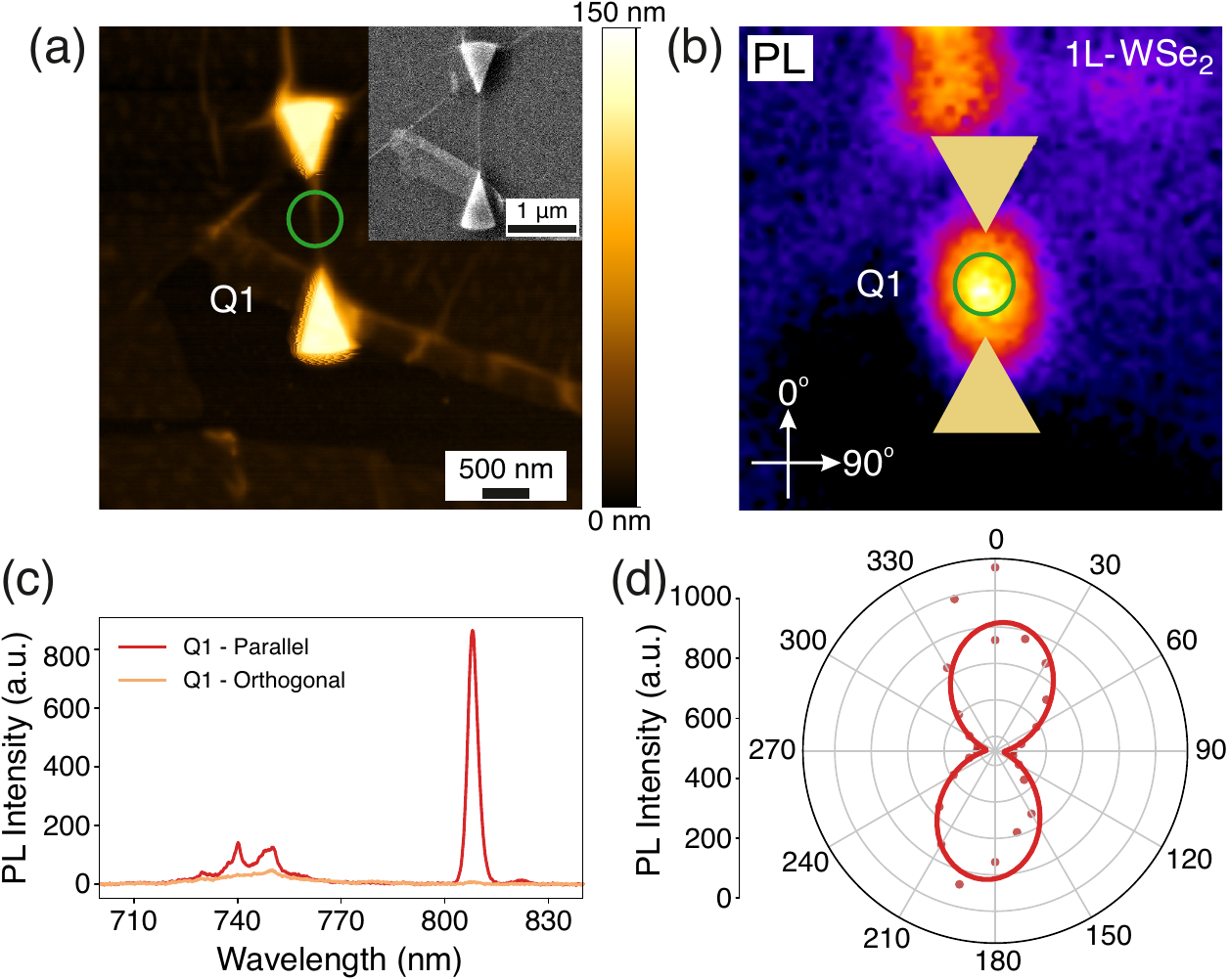}
\caption{\textbf{WSe\textsubscript{2} quantum emitters with bowtie (BT) shaped pillars.} (a) AFM image of a WSe\textsubscript{2} monolayer flake deposited on a bowtie pair of triangular nanostructures. A single nanowrinkle connects the two structures and is visible both in the main image and in the SEM image of the corresponding inset. (b) The corresponding PL image, taken at T = 4 K, exhibits increased photon emission along the nanowrinkle connecting the triangles. The two orthogonal axes are given as a reference for the upcoming polarization-resolved spectrum and polar plot. Both (a) and (b) highlight the collection spot (Q1) for the characterized emitter. (c) The presented PL spectra are measured from emitter Q1 under parallel and orthogonal in-plane polarization with respect to the direction of the nanowrinkle, which coincides with the two orthogonal axes given in (b). (d) Corresponding polarization-resolved polar plot of the PL intensities from emitter Q1. The sinusoidal fitting reveals a degree of linear polarization of 92 $\pm 12\%$.}
\label{Figure 4}
\end{figure}

In Figure \ref{Figure 3}, we present similar data for investigating the proposed FS design. The AFM image in Figure \ref{Figure 3}a showcases the surface morphology of the FS sample and the resulting nanowrinkles. Once more, the false coloring highlights the nanowrinkles, while the false-shading representation is shown in the inset for better contrast. Here, one of the five vertices has no visible nanowrinkle forming at its end. This is considered a normal occurrence due to the intricacies of the transfer process. Moreover, we observe that in the other four vertices (P1-P4), the nanowrinkles are indeed significantly shorter than those in the TS, leading to more localized SPEs. Despite this, the nanowrinkles are still well separated and can be addressed individually.\\
To further analyze the optical properties of the monolayer WSe$_2$ on the FS structure, we obtain the corresponding PL image (T = 4 K) under a 470 nm LED excitation, as shown in Figure \ref{Figure 3}b. The PL signal significantly increases around the tips where the nanowrinkles are located (P1-P4), with the tip missing a nanowrinkle showing emission comparable to the pristine monolayer.
To understand the effect of the nanowrinkles of the FS structure on the in-plane polarization of the single photons and compare it to that of the TS, we obtain polarization-resolved \textmu PL spectra from the four highlighted spots. We extract the PL intensities and present them as polar plots in Figures \ref{Figure 3}c-f. Similarly to Figure \ref{Figure 2}, the sinusoidal fitting reveals that the emitted photons from each SPE demonstrate high DOLP, with the extracted values at 86 $\pm 2\%$, 81 $\pm 2\%$, 76 $\pm 3\%$, and 85 $\pm 11\%$, respectively. More importantly, the results reaffirm that the in-plane polarization's orientation follows the nanowrinkles' directionality, although our control over this directionality is still unsatisfactory.
In general, the capability of the FS structure to host more localized and highly polarized emitters comes with the trade-off that certain vertices may not be able to form nanowrinkles.
\\
While both the TS and FS structures demonstrate some promising amount of control over both the SPEs' localization and polarization, it is worth restating that the control of nanowrinkle directionality is not flawless. As shown in Figures \ref{Figure 2} and \ref{Figure 3}, possible misalignments between the arm and the nanowrinkle are possible. This behavior is attributed to the hexagonal lattice structure of the material and how its intrinsic ``armchair" and ``zigzag" directions \cite{le2013joined} might affect the final directionality of the nanowrinkles. Moreover, a statistical analysis of the deviation in alignment between the directionality of the arms and the nanowrinkles of the TS and FS structures is presented in Figure S6 of the Supporting Information. The Gaussian fitting of the analysis reveals a central position of $0.25^\circ \pm 2.55^\circ$. However, the standard deviation of $16.0^\circ \pm 2.5^\circ$ highlights the demand for an improved design. Such structure (BT) is presented in Figure \ref{Figure 4}. The BT design facilitates the formation of the nanowrinkle between the two triangles, localizing the emitters to the best of our ability. Additionally, any potential misalignment between the vertices and the nanowrinkle would be minimized while isolating the SPEs from any naturally forming deformation in the surrounding area. Figure \ref{Figure 4}a shows the AFM image of an exemplary BT structure with a monolayer WSe$_2$ flake. A nanowrinkle connecting the two triangles is clearly visible in both the AFM image and the SEM inset. 
Figure \ref{Figure 4}b presents the corresponding PL image of the structure, showing increased photon emission in the nanowrinkle (Q1). Furthermore, increased photon emission is also observed in the left corner of the top triangle, which is attributed to a nanowrinkle formed along the edge of the structure. Nevertheless, the separation between the two regions is sufficient to characterize them individually. Figure \ref{Figure 4}c presents the polarization-resolved \textmu PL spectra, with the extracted intensities reported in the polar plot of Figure \ref{Figure 4}d. Once again, the orientation of the in-plane polarization agrees with the directionality of the nanowrinkle, with an extracted DOLP of 92 $\pm 12\%$.

In addition, Figure S7 presents SEM images of four different exemplary BT structures, all hosting a nanowrinkle connecting the two triangles, with a bar graph illustrating the general trend of directionality. Most BT nanowrinkles are aligned along within $\pm 2^\circ$ from the desired orientation. In contrast to the TS and FS designs, the BT seems to address all the roadblocks, achieving better alignment of the nanowrinkle while sufficiently localizing the emitters and isolating them from the surrounding environment. \\
After examining all suggested designs, we have demonstrated BT structures that effectively control the in-plane polarization of the emitted single photons while dictating the spatial localization of the SPEs.

\subsection{Characterization of single-photon emission}

\begin{figure*}[t!]
\includegraphics[width=0.7\textwidth]{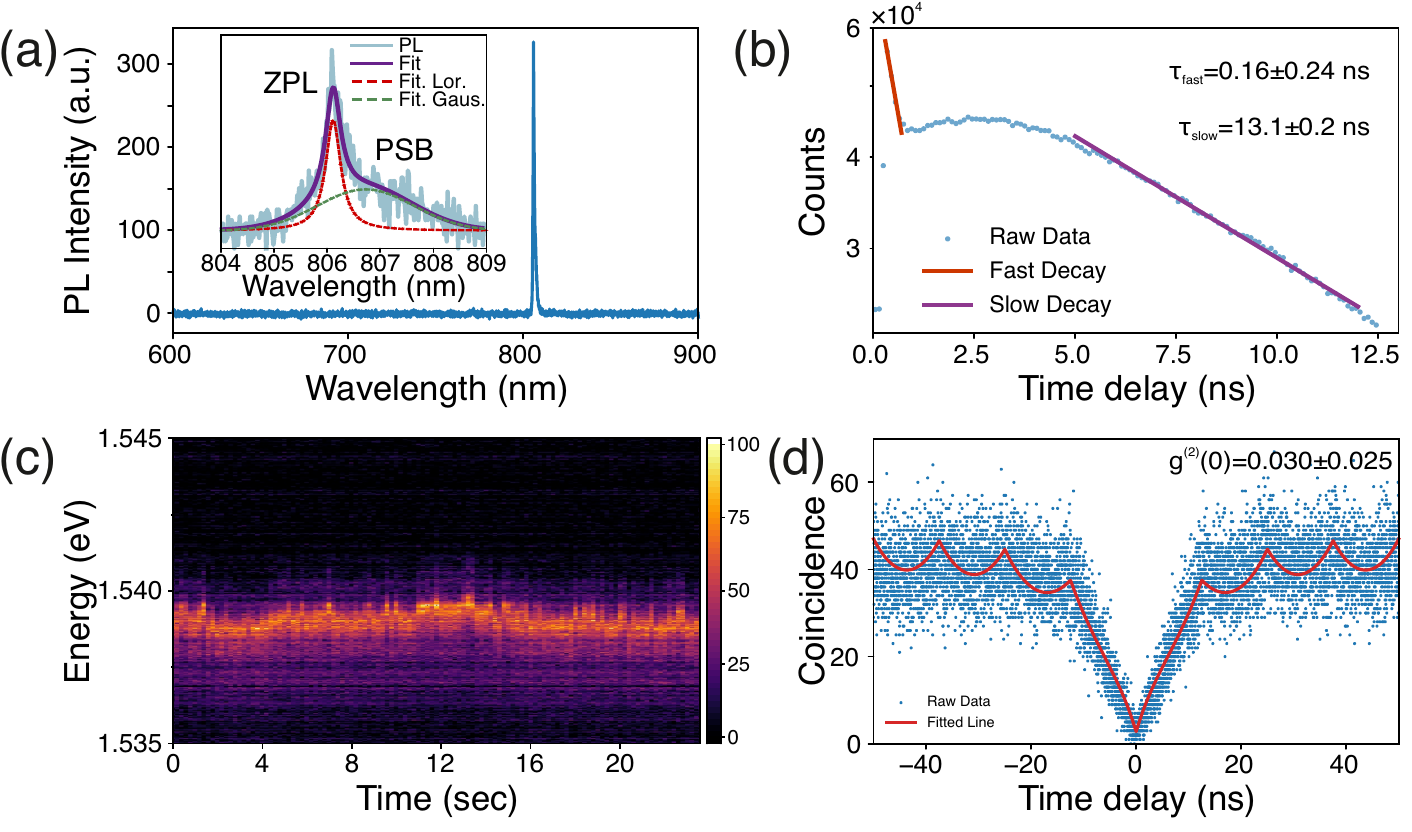}
\caption{\textbf{Single-photon characterization.} (a) PL spectrum (4 K) taken from an exemplary single localized quantum emitter with high-degree polarization of 92 $\pm 12\%$ under the above-band excitation of 532 nm femtosecond pulsed laser (80 MHz). The inset shows a high-resolution PL spectrum with a pronounced zero-phonon line (ZPL) and a low-energy broader phonon-side band. The fitting results exhibit the ZPL located at 806.1 nm, with a linewidth of $0.35 \pm 0.08$ nm. The broader PSB with a linewidth of $0.91 \pm 0.04$ nm is red-detuned by  0.6 nm to the ZPL, resulting in an asymmetric PL emission spectrum.
(b) Semi-logarithmic plot of the lifetime measurement of the WSe$_2$ quantum emitter fitted with a biexponential function, resulting in a lifetime of $\tau_\text{fast} = 0.16 \pm 0.24$ ns and $\tau_\text{slow} = 13.3 \pm 0.2$ ns.
(c) Time trace of PL emission signal, recorded using a high-resolution spectrometer with an integration time of 0.2 seconds per frame. The time trace data reveals a maximum wobbling span of approximately 1 meV for the ZPL. Fitting the distribution of the energy of the maxima for each line in the spectral map gives an average central position of the peak at 1.5365 eV with a standard deviation of 0.267 meV. (d) Corresponding second-order autocorrelation measurement ($g^{(2)}(\tau)$) of the emitter under the pulsed above band excitation, revealing an antibunching value of $g^{(2)}(0) = 0.030 \pm 0.025$. 
}
\label{Figure 5}
\end{figure*}

In order to assess the quality of the fabricated SPEs based on the novel structures presented in this work, we proceed to the optical characterization of an exemplary WSe$_2$ nanowrinkle quantum emitter. For \textmu PL measurements, we alter the excitation scheme from the CW LED to a 532 nm femtosecond pulsed laser (80 MHz) to optically excite the emitter. Supporting Note S3 and Figure S8 of the Supporting Information present the reasoning for that change. Figure \ref{Figure 5}a shows a broad range (600-900 nm) \textmu PL spectrum of an exemplary WSe$_2$ nanowrinkle emitter with a central wavelength of $\sim$806.1 nm. A 750 nm long-pass filter was used to spectrally isolate the single-photon emission line from the delocalized exciton emission and excitation laser. The inset of Figure \ref{Figure 5}a shows a high-resolution PL spectrum of the same emitter around its central wavelength, which exhibits a prominent zero-phonon line (ZPL) with a lower-energy phonon side band (PSB). The spectrum is fitted using a sum of a Lorentzian, representing the ZPL contribution (dashed red line), and a Gaussian, which captures the information related to the PSB contribution (dashed green line). We extract the linewidths of the two fitted curves, giving values of $0.35 \pm 0.08$ nm and $0.91 \pm 0.04$ nm, respectively, with the central wavelength of the PSB being red-detuned by 0.6 nm from the ZPL, resulting in an asymmetric PL emission spectrum. The increased contribution of the PSB to the total emission, evidenced by the ratio between the intensities ZPL : PSB = 49.2 \% : 50.8 \%, suggests a strong exciton-phonon coupling for WSe$_2$ monolayer SPEs. \cite{mitryakhin2023engineering} Even though the linewidth of the emission line is not Fourier-limited, being able to attain a narrow ZPL and resolve the PSB is attributed to the fabrication method that results in improved spatial isolation and a high linearly polarized emission. 
\\
To investigate the decay dynamics of the emitter, we perform the time-resolved PL (TRPL) measurement under the above-band pulsed excitation (Figure \ref{Figure 5}b). Our data reveal two distinct decays, which are fitted with two separate exponential functions to extract the corresponding decay rates. The origin of the initial fast decay ($\tau_\text{fast} = 0.16 \pm 0.24$ ns) is not clear at this stage; however, it seems to come from a rapid process, limited by the instrument response function (IRF). A more detailed analysis is presented in Supplementary Note S5 and Figure S10 of the Supporting Information. 
Nevertheless, despite its increased amplitude, the contribution of the fast decay accounts only for $\sim 6\%$ of the total curve. In contrast, the slow decay ($\tau_\text{slow} = 13.3 \pm 0.2$ ns) is the one that dominates the time-resolved response, making the largest contribution to the overall curve. Moreover, the calculated $\tau_\text{slow}$ is within the range of the reported lifetimes for localized SPEs in WSe$_2$ monolayer on dielectric substrates\cite{Kumar2015Strain-InducedWSe2, tripathi2018spontaneous, So2021PolarizationWSe2, Iff2021Purcell-EnhancedCavity, Branny2017DeterministicSemiconductor, he2015single, srivastava2015optically}. Finally, the abnormal increase in counts before the second decay begins may be caused by several factors. We attribute this rising behavior to trap states or excitation transfer processes that could lead to delayed decay and an initial rise in lifetime. Additionally, by potentially involving interactions with defects, non-radiative pathways may compete with the radiative decay, contributing to this rise.

While these complex decay dynamics require further investigation, their spectral stability is another critical aspect influencing the overall quality of SPEs.
To study the stability of the emission signal from the fabricated WSe$_2$ nanowrinkle, we obtain the high-resolution time trace of the PL emission with an integration time of 0.2 seconds per frame under the same above-band laser excitation scheme. We observe wobbling of the emission line, which could be attributed to the unstable charge environment and noise due to photo-excited free carriers under the above-band excitation. This unstable environment can be explained by the lack of encapsulation with a dielectric (e.g., hBN), which could minimize the interactions between the free charges and our SPE. For a qualitative analysis of this fluctuation, we perform a Gaussian fitting on the corresponding histogram presented in Figure S11, which reveals a fluctuation of the ZPL over $\pm$ 0.5 meV around the mean central position of 1.5365 eV.

To evaluate the single-photon purity from the fabricated emitters, we performed the second-order correlation measurement under the aforementioned 80 MHz pulsed excitation using a typical HBT setup (Figure \ref{Figure 5}d). The emission line from a representative nanowrinkle emitter was spectrally isolated by implementing a 750 nm long-pass filter. We observed peaks overlap at non-zero delays, separated by the laser repetition time of 12.5 ns, resulting from the increased lifetime of the emitter. More importantly, the significantly low coincidence counts at zero time delay indicate the high purity of single-photon emission from the sample. To quantitatively assess single-photon purity, we employed a double-sided exponential function fitting for the adjacent peaks, yielding a value of $g^{(2)}(0) = 0.030 \pm 0.025$. \\
The results presented in the previous sections demonstrate the effectiveness of the method proposed in this study. Our novel approach reliably produces high-quality single-photon emitters characterized by both high single-photon purity and controlled polarization.
Nevertheless, addressing the issues associated with previously discussed complicated dynamics and emission wobbling of our emitters is crucial for advancing photonic quantum information technology with TMD quantum emitters. These behaviors point to potential sources of additional inhomogeneous broadening, which can become the primary bottleneck for achieving lifetime-limited emission lines, resulting in poor indistinguishability.
Therefore, turning our attention to identifying and dealing with possible inhomogeneous broadening could be the next step toward developing TMD-based quantum emitters with high purity and indistinguishability.
Several strategies could be implemented to deal with such additional broadening.

On the one hand, it is essential to suppress the random charge fluctuations induced by free carriers on the substrate and the direct interaction of the emitters with the environment. One could implement a proper surface stabilization technique, such as encapsulation with Al$_2$O$_3$ via atomic layer deposition \cite{kim2016enhanced} or hBN encapsulation  \cite{Parto2021DefectK, kutrowska2022exploring} which can be directly implemented to the method presented in this work. An alternative approach involves the implementation of electrical contacts \cite{schadler2019electrical,zhai2020low}. On the other hand, an additional source of emission instability could result from the photo-induced charge carriers under the above-band optical excitation. One way to tackle this limitation lies in the employment of advanced excitation schemes, such as quasi-resonant \cite{bao2020probing}, resonant \cite{kumar2016resonant, somaschi2016near}, and SUPER excitation \cite{PRXQuantum.2.040354, Karli2022SUPEREmitter,vannucci2024single}.  Implementing these schemes \cite{bao2020probing, kumar2016resonant, PRXQuantum.2.040354, Karli2022SUPEREmitter} in TMD quantum emitters can facilitate the efficient exciton population and the mitigation of additional exciton-phonon interactions.

\section{Conclusion}

In this work, we presented an efficient method for generating highly polarized single photons from orientation-controlled nanowrinkles in monolayer WSe$_2$. This method employs novel nanostructures to induce one-directional strain deterministically, enabling efficient control of the polarization of emitted single photons. We have successfully demonstrated single photon emission from WSe$_2$ nanowrinkle emitters with high purity ($g^{(2)}(0) = 0.030 \pm 0.025$) and a high degree of polarization (99 $\pm$ 4\%). These findings provide crucial insights for the future integration of these quantum emitters into a specific mode of desired photonic structures, such as the TE mode of a planar ridge waveguide. Furthermore, understanding the fabrication limitations and critical aspects of deterministic polarization control of quantum emitters are essential steps toward their effective utilization as functional single-photon sources in quantum photonics.

\section{Methods} \label{section:methods}

\subsection{Sample preparation}

\textit{Nanopillar fabrication:} The substrate was prepared by thermally oxidizing 110 nm of SiO$_2$ on a silicon wafer. Alignment marks were patterned by UV lithography (using the AZ 5214E positive resist) and subsequent e-beam evaporation of 5 nm of Ti and 50 nm of Au. 
After cleaving the wafer into chips, the fabrication of the nanopillars was carried out by spin coating a high-resolution negative e-beam resist (hydrogen silsesquioxane or HSQ, XR-1541-006) at 3000 rpm for 1 min, followed by two soft baking steps at 120$^\circ$C and 220$^\circ$C, both for 2 minutes. The resist was patterned by an e-beam writer (JEOL 9500, 100 kV, 6 nA) with a dose of 11000 \textmu C/cm$^2$ and developed in a 1:3 solution of AZ 400K : H$_2$O to obtain 150 nm-tall glass nanopillars in four different shapes, namely cylindrical pillars, three-pointed stars, five-pointed stars, and bowties. 
It is worth remarking that a high dose is required to cure the spin-on dielectric, whose composition becomes SiO$_2$. Other negative e-beam resists with high sensitivity (activated at doses as low as 30 \textmu C/cm$^2$) have also been tested (Supporting Note S5).   

\textit{Exfoliation and transferring:} WSe\textsubscript{2} flakes were first exfoliated from bulk crystals (HQ Graphene) by using the scotch-tape method \cite{castellanos2014deterministic}. The flakes were then exfoliated on a polydimethylsiloxane (PDMS) stamp, which was prepared on a glass slide. The monolayers were identified through their PL emission at RT by using a 450 nm LED source and were then dry-transferred on the nanostructures with a transfer stage, which allows heating the chip to 70$^\circ$C. At this temperature, the van der Waals interactions between the flake and the substrate overcome the adhesion to PDMS, and the flake is released.

\textit{Defect fabrication:} Defects were introduced on the monolayer lattice by bombarding the strained areas of the material with an e-beam (JEOL 9500 e-beam writer, 100 kV, 1000 \textmu C/cm$^2$).

\subsection{Optical characterization} 
\textit{Photoluminescence measurements:} The PL measurements were obtained with a custom-built low-temperature micro-photoluminescence (\textmu PL) setup. The sample is mounted on a closed-cycle cryostat (attoDRY800) operating at a base temperature of 4 K. The cryostat is equipped with piezoelectric nanopositioners and a low-temperature microscope objective ($60 \times$ NA = 0.82) located inside the cryostat. The sample was excited with a CW LED at 470 nm for the PL imaging. For the microphotoluminescence spectroscopy,  a pulsed (80 MHz) femtosecond laser (Chameleon Ultra II, Coherent) at 532 nm was employed for the optical excitation of the emitters. Further investigation of the effect of the different excitation schemes on the emitters is presented in Supporting Note 3 and Figure S8 of the Supporting Information.

The PL spectra were acquired with a fiber-coupled spectrometer (iHR 550, Horiba). A polarizer mounted on a motorized rotation mount was placed on the collection path for polarization-resolved measurements. PL images were acquired with an integration time of 1 s using a CMOS camera (Tucsen Dhyana, 4 MP). 

\textit{Second-order correlation measurements:} For second-order correlation measurements, we employed a fiber-optic Hanbury-Brown and Twiss (HBT) interferometer. The photon counting was conducted using superconducting nanowires single-photon detectors (ID218, ID Quantique) connected to a time controller (ID900, IDQ).

\subsection{Sample imaging} 
Bright-field (BF) images of the sample were taken using a confocal-microscopy setup equipped with an objective lens (Nikon LU Plan Fluor, $50 \times$, NA = 0.8) and a CMOS camera (CS165MU/M, Thorlabs). \\
The topographic images of the samples were obtained with a scanning electron microscope (SEM) (Zeiss Supra 40 VP, 5 kV, SE2 detector). To avoid creating further defects in the flakes, SEM imaging was carried out after all the measurements for optical characterization. \\
Finally, the surface morphology of the samples, namely the width and height of the nanopillars and the WSe$_2$ nanowrinkles, was characterized by atomic force microscopy (Bruker Dimension Icon-PT, ScanAsyst-air tip).

\section*{Associated Content}
\noindent
\textbf{Supporting Information}

\noindent
The Supporting Information is available free of charge at \url{https://xxxxxxxx.xxxxx.}

Figure S1: Schematic representation of the fabrication flow, Figure S2: Comparison of PL spectra before and after e-beam irradiation, Figure S3-5: Investigation on the morphological profile of the nanowrinkles for all different structures, Figure S6: SEM images of TS and FS structures with a statistical analysis on the misalignment of the nanowrinkles, Figure S7: SEM images of BT structures with a histogram of misalignment of the nanowrinkles, Figure S8: Comparison of the PL emission of an exemplary quantum emitter under two excitation schemes (LED, 470 nm and pulsed laser, 532 nm), Figure S9: AFM image of a WSe$_2$ monolayer flake on a TS structure made from AR-N 7520.11., Figure S10: Comparison between the lifetime of the delocalized exciton and the exemplary SPE studies in Figure 5, Figure S11: Histogram regarding the spectral wobbling of the ZPL studied in Figure 5 \\

\section*{Funding}

The authors acknowledge support from the European Research Council (ERC-StG ``TuneTMD", grant no. 101076437) and the Villum Foundation (grant no. VIL53033). The authors also acknowledge the European Research Council (ERC-CoG ``Unity", grant no. 865230).

\section*{Notes}
The authors declare no competing financial interest.

\section*{Acknowledgements}
The authors acknowledge the cleanroom facilities at DTU Nanolab – National Centre for Nano Fabrication and Characterization.  

\section*{Author contributions}

AP fabricated the quantum emitter samples. CP and AM supported the project by fabricating various shaped pillars for strain engineering. AP, CP, AM, and BM performed optical measurements from the fabricated samples. AP, PM, and CP contributed to data analysis and processing. BM conceived the idea, and LV, NG, and BM coordinated the project. AP, PM, and BM wrote the manuscript with input from all co-authors.\\

\section{References}
\bibliography{ref} 

\end{document}